\documentclass[10pt,twocolumn,twoside]{IEEEtran}

\usepackage{flushend}

\usepackage[utf8x]{inputenc}

\usepackage{amsmath,amssymb,mathtools,amsthm}%
\usepackage{graphicx}
\usepackage{subfig}
\usepackage{color}
\usepackage{url}
\usepackage{caption}
\usepackage{hyperref}

\usepackage{times}

\usepackage[norelsize,ruled,vlined,linesnumbered]{algorithm2e}
\SetInd{0.2em}{0.7em}
\SetCommentSty{textit}
\SetAlgoInsideSkip{smallskip}
\DontPrintSemicolon

\usepackage{Cris}

\renewcommand{\blue}[1]{#1}
\renewcommand{\green}[1]{#1}

\newcommand\AFc[1]{}
\newcommand\AFs[1]{}

\newcommand\removed[1]{}

\def\Boldd{{\mathbf{d}}}
\def\Bolde{{\mathbf{e}}}

\def\Boldp{{{\mathbf{p}}}}

\def\Boldu{{\mathbf{u}}}
\def\Boldv{{\mathbf{v}}}
\def\Boldx{{\boldsymbol x}}

\def\Boldw{{\boldsymbol w}}

\def\Boldzero{{\boldsymbol 0}}

\def\Boldgamma{{\boldsymbol \gamma}}

\def\BoldA{{\mathbf{A}}}
\def\BoldD{{\mathbf{D}}}
\def\BoldF{{\mathbf{F}}}
\def\BoldG{{\mathbf{G}}}
\def\BoldI{{\mathbf{I}}}
\def\BoldL{{\mathbf{L}}}
\def\BoldM{{\mathbf{M}}}
\def\BoldP{{\mathbf{P}}}
\def\BoldQ{{\mathbf{Q}}}
\def\BoldS{{\mathbf{S}}}
\def\BoldX{{\mathbf{X}}}
\def\BoldY{{\mathbf{Y}}}

\def\Boldone{{\bf 1}}
\newcommand*\Bell{\ensuremath{\boldsymbol\ell}}

\begin{document}

\title{Online Leader Selection for Improved\\ 
Collective Tracking and Formation Maintenance} %

\author{
Antonio Franchi$^{1,2}$ and Paolo Robuffo Giordano$^{3}$%
\thanks{\hspace{-1em}$^1$CNRS, LAAS, 7 avenue du colonel Roche, F-31400 Toulouse, France } 
\thanks{\hspace{-1em}$^2$Univ de Toulouse, LAAS, F-31400 Toulouse, France {\tt \scriptsize \href{mailto:afranchi@laas.fr}{afranchi@laas.fr}}}
\thanks{\hspace{-1em}$^3$CNRS at Irisa and Inria Rennes Bretagne Atlantique, Campus de Beaulieu, 35042 Rennes Cedex, France. E-mail: {\tt \scriptsize  prg@irisa.fr}}
}

\maketitle

\begin{abstract}
The goal of this work is to propose an extension of the popular leader-follower framework for multi-agent collective tracking and formation maintenance in presence of a time-varying leader. In particular, the leader is persistently selected \emph{online} so as to optimize the tracking performance of an exogenous collective velocity command while also maintaining a desired formation via a (possibly time-varying) communication-graph topology.
The effects of a change in the leader identity are theoretically analyzed and exploited \blue{for defining} a suitable error metric able to
capture the tracking performance of the multi-agent group.
Both the group performance and the metric design are found to depend upon the spectral properties of a special \emph{directed graph} induced by the identity of the chosen leader. %
By exploiting these results, as well as distributed estimation techniques, %
we are then able to detail a fully-decentralized adaptive strategy able to periodically select \emph{online} the \emph{best leader} among the neighbors of the current leader. Numerical simulations show that the application of the proposed technique results in an improvement of the overall performance of the group behavior w.r.t.~other possible strategies.
\end{abstract}

\IEEEpeerreviewmaketitle

\vspace{5pt}

\section{Introduction}

\IEEEPARstart{M}{any} complex organisms made of several entities rely on the basic property of being able to follow an external source. This is for example the case of groups of animals during pack-hunting of a prey, or migrations driven by natural signals.
Inspired by these considerations, several collective tracking behaviors and control algorithms have been proposed for multi-agent systems~\cite{2006-Olf,2008-RenBea,2010-RenCao} as, for instance, the well-known leader-follower paradigm, one of the most popular techniques in the control and robotics communities~\cite{2001-LeoFio,2009-MarMorPraVanMicPapDan,2010-GusDimEgeHu_,2010-MorMarPra,2010-CheSunYanChe,2010-TwuEgeMar,2011-NotEgeHaq}.
In the leader-follower scenario, a special agent (the leader) has access to the signal source, e.g., to the reference motion to be tracked by the whole group. In order to act cooperatively, this local information must then be spread among the rest of the group by means of proper local actions (see, e.g.,~\cite{2013-Ant} where distributed formation control and leader-follower approaches are thoroughly reviewed).

Within the leader-follower scenario, one of the main research topics has been the study of new distributed estimation and control laws able to \emph{i)} propagate the reference motion signal through local communication to the whole group and \emph{ii)} let the group track this reference with the smallest possible error/delay.
In most of the cases, however, the leader is assumed to be a particular (constant) member chosen by the group at the beginning of the task. 
This problem, denoted here as \emph{static leader election}, has been deeply investigated for autonomous multi-agent systems. In the static leader election case, the problem is to find a distributed control protocol such that, eventually, one (and only one) agent takes the decision of being the leader~\cite{1997-Lyn}.
Among other works, in~\cite{2009-BulCorMar} the leader election problem is solved by the {\tt FLOODMAX} distributed algorithm using explicit message passing among the formation. In~\cite{2010-ShaTeiSanJoh}, the leader election problem is solved using fault detection techniques and without explicit communication, as done by some animal species. However, in all these works the leader election is assumed to be performed \emph{only once}, e.g., at the beginning of the task, with the goal of selecting a suitable leader whose identity is then retained for the whole mission duration. %

On the contrary, in this paper we extend this paradigm by assuming that \emph{i)} the identity of the leader is an additional degree of freedom that can be \emph{persistently changed} (i.e., online) with the aim of \emph{ii)} %
optimizing both the group tracking performance of the reference motion command and the (concurrent) convergence to a desired group formation. We refer to this problem as \emph{online leader selection}.

In the recent years, a few works have addressed related objectives with different approaches. Maximization of network coherence, i.e., the ability of the consensus-network to reject stochastic disturbances, has been the optimization criteria used in~\cite{2010-PatBam}. 
The criteria used in \cite{2010-BorAtt} \blue{have} been controllability of the network and minimization of a quadratic cost to reach a given target.
The case of large-scale network and noise-corrupted leaders has been considered in \cite{2011-FarLinJov,2011-LinFarJov}. 
A joint consideration of controllability and performance has been recently considered in~\cite{2012-ClaBusPoo}.
The authors in~\cite{2012-KawEge} use instead the concept of manipulability to select the best leader in the group.

{With respect to these cases, we consider a different optimization criterion which, we believe, is more suited for applications involving collective motion tracking: the convergence rate to the reference velocity signal (only known by the current leader) and to the desired formation. We note that these criteria do not only depend on the characteristics of the network, but also on the current state of the agents and on the current reference signal. Therefore, their optimization cannot be performed once and for all at the beginning of the task, as it is the case for most of the aforementioned approaches.}
Furthermore, for the sake of generality, we also consider the possibility of a \emph{time-varying} (but connected) interaction graph, and we provide a \emph{fully-distributed} control strategy for obtaining an optimal and online selection of the leader.

The main contributions of this work can be summarized as follows: we introduce a new leader-follower paradigm in which the agents can persistently change the current leadership in order to adapt to both the variation of an external signal source (to be tracked by the group), and to a possibly time-varying communication-graph topology. \blue{For what concerns the} motion tracking algorithms, we consider 
\blue{a widely used consensus-like decentralized multi-agent coordination model (see, e.g.,~\cite{2010-MesEge})} and theoretically analyze the effects of a changing leadership over time. This is obtained by proposing a suitable error metric able to quantify the performance of the multi-agent group in tracking the external reference signal and in achieving the desired formation shape. We then propose a fully-decentralized \emph{online leader selection algorithm} able to periodically select the `best leader' among the neighbors of the current leader, and we finally provide numerical simulations to show the effectiveness of our approach. %
A preliminary version of the framework proposed in this paper has been presented in~\cite{2011e-FraBueRob}, where, however, simpler metrics have been considered,  formal proofs were \blue{omitted}, and simpler case studies were discussed.

The paper is organized as follows.  Section~\ref{sec:coll_tracking} defines the problem background and introduces some preliminary results. Section~\ref{sec:lead_sel_first} presents the first main contribution of the paper by theoretically analyzing the effect of a changing leader on the considered tracking performance. Section~\ref{sec:decentr_algo} presents the second main result of the paper by proving tha the selection of the best leader can be performed in a completely decentralized way.
Finally, sections~\ref{sec:simul} and~\ref{sec:concl} present some numerical examples on the theoretical results and a final discussion, respectively.  

\section{Modeling Of Formation Maintenance And Tracking Of An External Reference}\label{sec:coll_tracking}  %

This section introduces the general model of our multi-agent scenario and the first contribution of our paper, i.e., a set of results concerning the considered general model. We consider a group of $N$ mobile agents modeled as points in $\mathbb{R}^d$, with  $d\in\{2,3\}$, whose positions are denoted with $\Boldp_i\in\mathbb{R}^d$ for $i=1\ldots N$.  As customary, we model the inter-agent communication capabilities by means of the (symmetric) \emph{adjacency matrix} $\BoldA=\{A_{ij}\}\in\{0,1\}^{N\times N}$ with $A_{ij}=1$ if agents $i$ and $j$, $j\neq i$, can communicate, and $A_{ij}=0$ otherwise, $\forall\; i,j=1\ldots N$.
We also denote with  $\mathcal{N}_i = \{j \,|\, A_{ij} = 1\}$ the set of \emph{neighbors} of agent $i$, i.e., the agents with which $i$ can communicate, and let $\mathcal{G}$ represent the \emph{undirected communication graph} defined by the adjacency matrix $\BoldA$. 
Finally, we denote with $\BoldL$ the Laplacian matrix of $\calG$, i.e., $\BoldL={\rm diag}(\BoldA\Boldone)-\BoldA$, where $\Boldone$ represents a column vector of all ones of proper size ($N$, in this case), and ${\rm diag}$ returns the diagonal matrix associated to a vector.
We assume that $\calG$ is connected,
 i.e., there
 exists a sequence of hops (edges) connecting any pair of agents in the communication network\footnote{One can always restrict the analysis to a suitable connected component of the group.}. 
As well known, see, e.g.,~\cite{2010-MesEge}, this implies that $\BoldL$ has rank $N-1$,
or, equivalently, that the second smallest eigenvalue $\lambda_2$ of $\BoldL$ (the algebraic connectivity of $\calG$) is positive.

\begin{figure}
\includegraphics[width=0.99\columnwidth]{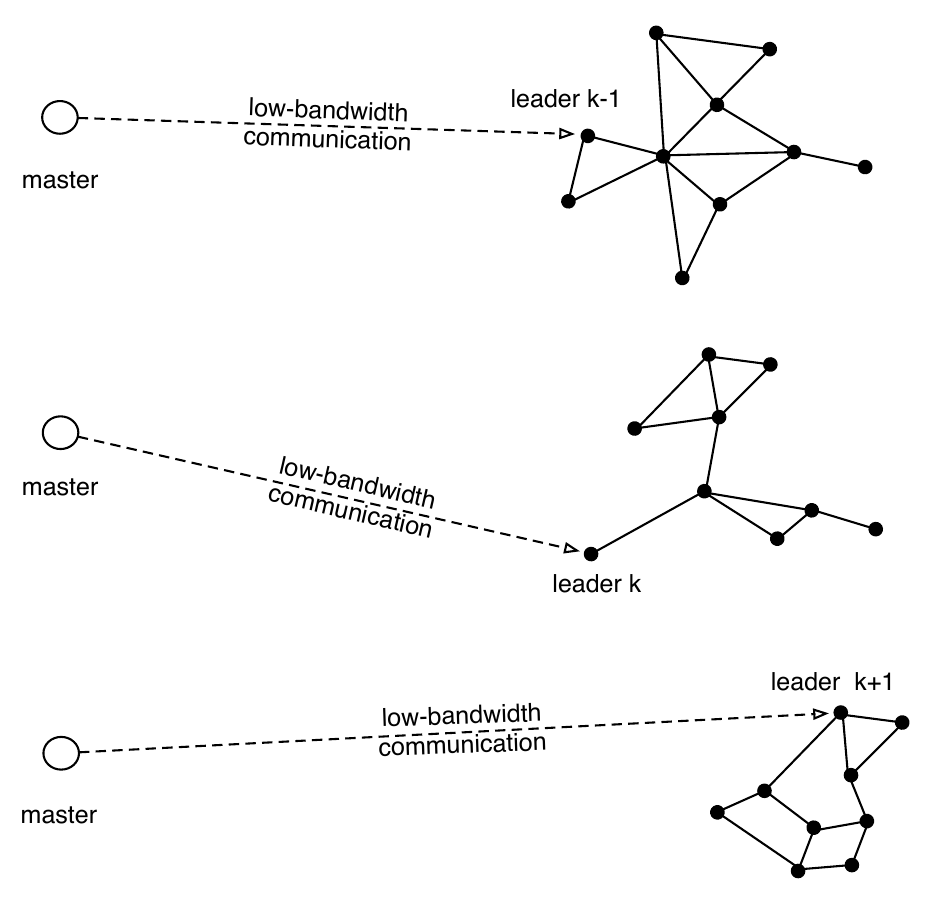}
\caption{\blue{Abstraction of the application scenario, in which a master agent (e.g., a base station) can only communicate with one an agent at the time (called leader) with a low bandwidth. The leader can be changed every time a new high-level command  from the master is sent to the group.}}
\label{fig:application}
\end{figure}

An `external entity', referred to as the \emph{master}\footnote{This nomenclature is borrowed from the teleoperation literature.}, provides a collective motion command  to the group in the form of a velocity \blue{reference} $\Boldu_r\in\mathbb{R}^{d}$. %
\begin{remark}\label{rem:UUVexample}
\blue{The group of agents may represent, e.g., a group of remote unmanned vehicles that needs to keep a fixed formation in order to monitor a given area, and the master can represent a base station in charge of guiding the group based on some additional (locally available) knowledge and computational power. In this situation, because of typical bandwidth limitations, especially over large distances, it is meaningful to assume that the master can only communicate with one particular agent in the group at the time, see, e.g.,~\cite{2012f-FraSecRylBueRob,2012h-RieFraRobBueSon} and references therein. Similarly, because of the same reasons, it is also meaningful to assume that the high-level command sent by the base station (the master) has a low frequency compared to the group internal dynamics. Therefore the agents will need to control their internal motion (faster dynamics) by `interpolating' between two consecutive high-level commands from the base station (e.g., by considering \emph{piece-wise constant} reference commands among consecutive receiving times).}\\
Figure~\ref{fig:application} provides a pictorial representation of the aforementioned application scenario.
\end{remark}

Because of the practical limitations discussed in Remark~\ref{rem:UUVexample} (which may arise in several different operating contexts), we then assume at this modeling stage that the master can communicate, \blue{with negligible delay,} the current value of $\Boldu_r\in\mathbb{R}^{d}$ to \emph{only} one agent at a time, called \emph{leader} from now on, and denoted with the index $l$ throughout the rest of the paper. 
\blue{We do not pose any special constraint on the identity of the initial leader. Furthermore, we assume that the master sends $\Boldu_r$ to the current leader at a known frequency $1/T_r$, with $T_r\geq 0$ being the sending period ($\Boldu_r$ will then be treated by the current leader as a \emph{constant vector} among consecutive receiving times). Symmetrically, the group can inform the master on the identity of the current leader at the same frequency $1/T_r$.}

Exploiting the multi-agent communication network, the reference velocity $\Boldu_r\in\mathbb{R}^{d}$ \blue{(only known to the leader)} can be however transmitted to the other agents of the group via a multi-hop propagation algorithm. As representative of the several existing possibilities in this sense, %
we consider here the following consensus-like law for easily modeling fast/slow propagation algorithms and technologies:
\begin{align}
\dot{\hat\Boldu}_i &= -k_u\sum_{j\in \mathcal{N}_i}(\hat\Boldu_i - \hat\Boldu_j) &\forall i\neq l \label{eq:vel_estimator} \\
\hat\Boldu_l&=\Boldu_r & \label{eq:vel_estimator_reset}
\end{align}
where $\hat\Boldu_i$ is the $i$-th estimation of $\Boldu_r$, and $k_u$ a positive scalar gain. Model~\eqref{eq:vel_estimator}--\eqref{eq:vel_estimator_reset} may approximate a large variety of propagation algorithms with different convergence speeds by simply tuning the gain $k_u$ (larger gains correspond to shorter propagation times and vice-versa). For example a  ultrasonic underwater communication can be modeled choosing a relatively `small' $k_u$ while a high-bandwidth LAN network should more reasonably be modeled with a larger $k_u$.

Letting $\hat\Boldu = (\hat\Boldu_1^T\ldots\hat\Boldu_N^T)^T\in\mathbb{R}^{Nd}$,~\eqref{eq:vel_estimator}--\eqref{eq:vel_estimator_reset} can be compactly rewritten as %
\begin{align}
\dot{\hat\Boldu} = -k_u(\BoldL_l \otimes \BoldI_d)\hat\Boldu = k_u\BoldG_l\hat\Boldu,\label{eq:est_ref_vel}
\end{align}
where $\BoldL_l$ is the `in-degree' Laplacian matrix of the directed graph (digraph) $\calG_l$ obtained from $\calG$ by removing all the in-edges of $l$, $\otimes$ is the Kronecker product, $\BoldI_d$ the $d\times d$ identity matrix, and $\BoldG_l=-(\BoldL_l\otimes \BoldI_d)\in\mathbb{R}^{Nd\times Nd}$. 
Using~\eqref{eq:est_ref_vel}, the velocity estimation error
\begin{align}
\Bolde_{\hat\Boldu} &= \hat{\Boldu} - \Boldone \otimes \Boldu_r\label{eq:error_u_def}\end{align}
obeys the dynamics
\begin{align}
\dot\Bolde_{\hat\Boldu} = k_u\BoldG_l\Bolde_{\hat\Boldu} - {\Boldone}\otimes\dot\Boldu_r.\label{eq:error_u_dyn}
\end{align}

We further assume that\blue{, besides collectively tracking the reference velocity $\Boldu_r$,} the agents must also arrange in space according to a desired formation defined in terms of a set of constant relative positions  taken as reference shape \blue{in some common frame decided before the task execution}. \blue{These relative positions are assumed generated as all the possible differences between pairs of positions in a set of $N$ absolute positions $\Boldd=(\Boldd_1^T\ldots\Boldd_N^T)^T\in\mathbb{R}^{dN}$. The `virtual' absolute positions $\Boldd$ are clearly defined `up to an arbitrary translation', since only the position differences will play a role for the coordination law.}

Such a formation control task is a typical requirement in many multi-agent applications (see, again, Remark~\ref{rem:UUVexample} for an example). A number of different control strategies can be employed to achieve this goal, depending on the actuation and sensing capabilities of the agents, \blue{see, e.g.,~\cite{2013-Ant} and references therein for the centralized task-priority framework, or~\cite{2010-MesEge} for the decentralized graph-theoretical methods}. 
In order to model a generic control action for letting the agents achieving the desired formation, we consider the classical and well-known \emph{distributed} consensus-like formation control law
\begin{align}
 \dot\Boldp_i &=
 \left\{{
\begin{matrix*}[l]
\hat\Boldu_i - k_p\displaystyle{\sum_{j\in \mathcal{N}_i}}((\Boldp_i-\Boldp_j) - (\Boldd_i - \Boldd_j))
 & i \neq l\\
 \hat\Boldu_i (= \Boldu_r), & i = l
\end{matrix*}}
\right.
\label{eq:EgerFixed}
\end{align}
where $\Boldd_i - \Boldd_j\in \mathbb{R}^3$ represents the desired relative position between neighboring agents $i$ and $j$, and $k_p>0$ is a positive scalar gain.
The complete agent dynamics then takes the form
\begin{align}
\dot{\Boldp} = \hat\Boldu + k_p\BoldG_l(\Boldp - \Boldd), \label{eq:whole_agent_dyn}
\end{align} 
\blue{
where $\Boldp = (\Boldp_1^T\ldots\Boldp_N^T)^T\in\mathbb{R}^{Nd}$.
}
The simple linear dynamics~\eqref{eq:whole_agent_dyn} is expressive enough for suitably modeling a generic (also non-linear) formation control action around its equilibrium point. The gain $k_p$ determines the `stiffness' of the formation control, i.e., how strongly the agents will react to deviations from their desired formation.

Letting $\Boldv=\dot \Boldp$, we now consider the following formation tracking error vector
\begin{align}
\Bolde_\Boldp = (\Boldp - \Boldone\otimes \Boldp_l) - (\Boldd - \Boldone\otimes \Boldd_l)\label{eq:error_p}
\end{align}
and velocity tracking error vector
\begin{align}
\Bolde_\Boldv &= \Boldv - \Boldone \otimes \Boldv_l = \Boldv - \Boldone \otimes \Boldu_r,\label{eq:error_v}
\end{align}
representing, respectively, the tracking accuracy of the desired formation encoded by $\Boldd$, and of the reference velocity $\Boldu_r$ \blue{(known by the current leader, and propagated to the other agents via~\eqref{eq:vel_estimator}--\eqref{eq:vel_estimator_reset})}.

Using the properties $\BoldG_{l}(\Boldp-\Boldd)=\BoldG_{l}\Bolde_{\Boldp}$, $\BoldG_{l}\Boldv=\BoldG_{l}\Bolde_{\Boldv}$, $\BoldG_{l}\hat\Boldu=\BoldG_{l}\Bolde_{\hat\Boldu}$, and taking into account~(\ref{eq:error_u_dyn})--\eqref{eq:whole_agent_dyn}, the dynamics of the overall error vector $\Bolde = (\Bolde_{\Boldp}^T \; \Bolde_{\Boldv}^T \; \Bolde_{\hat\Boldu}^T)^T$ then takes the expression
\begin{align}%
\dot\Bolde
=
\begin{pmatrix}
k_p\BoldG_{l} & \Boldzero_{Nd} & \BoldI_{Nd}\\
\Boldzero_{Nd} & k_p\BoldG_{l} & k_u\BoldG_l \\
\Boldzero_{Nd} & \Boldzero_{Nd} & k_u\BoldG_l
\end{pmatrix}
\Bolde
-
\begin{pmatrix}
\Boldzero\\
\Boldone\otimes\dot\Boldu_r\\
\Boldone\otimes\dot\Boldu_r
\end{pmatrix}.
\label{eq:err_p_v_dyn}
\end{align}%
\AFc{qui un reviewer dice: the velocity
reference signal $u_r$ is assumed to be a staircase signal
and therefore its derivative is not necessary. system (11) for
the error vector is actually an autonomous system from the beginning.}

\AFc{sottolineare che online per noi significa che possiamo basarci solo sul passato e presente non sul futuro. }

\blue{
As expected, the formulation~\eqref{eq:err_p_v_dyn} is quite general and, in fact, it has been exploited several times (in different contexts) in the multi-agent literature as, e.g., in~\cite{2013-Oh_Ahn}, where the same formulation is used for, however, other purposes not related to the leader selection problem considered in this work.}

We now show some fundamental properties of system~\eqref{eq:err_p_v_dyn} and of other relevant quantities instrumental for illustrating the main results of the paper. %
First of all let us rewrite matrix $\BoldL_l$, obtained from $\BoldL$ by zeroing its $l$-th row, as follows:
\begin{align}
\BoldL_l \triangleq 
\begin{pmatrix}
\BoldM_{l,1}       & \ell_{l,1} & \BoldM_{l,2}\\
{\bf 0}^T & 0 & {\bf 0}^T\\
\BoldM_{l,3}       & \ell_{l,2} & \BoldM_{l,4}\\
\end{pmatrix},\label{eq:L_partition}
\end{align}
where $\BoldM_{l,1}$, $\BoldM_{l,2}$, $\BoldM_{l,3}$, $\BoldM_{l,4}$, $\ell_{l,1}$, $\ell_{l,2}$, and ${\bf 0}$ are matrices and column vectors of proper dimensions. We also define 
$$
\BoldM_l \triangleq \begin{pmatrix}
\BoldM_{l,1} & \BoldM_{l,2}\\
\BoldM_{l,3} & \BoldM_{l,4}\\
\end{pmatrix}\in\mathbb{R}^{N-1\times N-1}
$$
and ${\Bell}_l \triangleq (\ell_{l,1}^T\, \ell_{l,2}^T)^T\in\mathbb{R}^{N-1}$.  
The following properties play a central role in the next developments.

\begin{property}\label{property:Ml}
Denoting with $\sigma(\BoldS)$ the spectrum of a square matrix $\BoldS$, and assuming connectedness of the graph $\calG$, the following properties hold:
\begin{enumerate}
\item $\BoldL_l \Boldone = {\bf 0}$, $\forall l=1\ldots N$;

\item $\BoldM_l\Boldone = (\Boldone^T \BoldM_l)^T ={\Bell}_l$;

\item $\BoldM_l$ is symmetric and positive definite;

 \item $\sigma(\BoldL_l) = \sigma(\BoldM_l) \cup \{0\}$.
\end{enumerate}
\end{property}
\begin{proof}
\AFc{Un reviewer dice: the expression of G in the proof. The notation used here is unclear and needs to be clarified by the authors.}
The first item follows from $\BoldL\Boldone = {\bf 0}$ which holds by construction, while the second item is a direct consequence of the first one.

\blue{
In order to prove the third item, consider the decomposition $\BoldM_l = \BoldL_{-l} - {\rm diag}({\Bell}_l)$, where $\BoldL_{-l}\in\mathbb{R}^{{N-1}\times{N-1}}$ is the Laplacian of the subgraph $\calG_{-l}$  obtained from $\calG$ by removing the $l$-th vertex (and all its adjacent edges), and $-{\rm diag}({\Bell}_l)\in\mathbb{R}^{{N-1}\times{N-1}}$ is a  diagonal matrix built on top of vector ${\Bell}_l$, i.e., with `ones' in all the diagonal entries corresponding to the vertexes of $\calG_{-l}$ adjacent to $l$ in $\calG$ and `zeros' otherwise.

Both matrix $-{\rm diag}({\Bell}_l)$ are $\BoldL_{-l}$
are positive semidefinite. In fact, the eigenvalues of $-{\rm diag}({\Bell}_l)$ are either $1$ or $0$ by construction, while $\BoldL_{-l}$ is the Laplacian matrix of a graph, which is always positive semidefinite~\cite{2010-MesEge}.
Therefore $\BoldM_l$ is at least positive semidefinite, being the sum of two positive semidefinite matrixes. We prove now that $\BoldM_l$ is actually positive definite by showing that $\forall\Boldw\in\mathbb{R}^{N-1}$, $\Boldw\neq \boldsymbol{0}$, we have that $\Boldw^T\BoldM_l\Boldw>0$.
Exploiting the aforementioned decomposition we obtain
\[
\underbrace{\Boldw^T\BoldM_l\Boldw}_{=b_1+b_2} = \underbrace{\Boldw^T\BoldL_{-l}\Boldw}_{=b_1\ge 0} + \underbrace{\Boldw^T(-{\rm diag}({\Bell}_l))\Boldw}_{=b_2\ge0}.   
\]
We now prove now that $\forall\Boldw\in\mathbb{R}^{N-1}$, $\Boldw\neq \boldsymbol{0}$, $b_1=0\Rightarrow b_2>0$ which in turns will imply that $\forall \Boldw\in\mathbb{R}^{N-1}$ $b_1+b_2>0$, i.e., that $\BoldM_l$ is positive definite.

 From the properties of a Laplacian matrix, the subspace of vectors $\Boldw$ such that $\Boldw^T\BoldL_{-l}\Boldw=0$ is spanned by the eigenvectors $\Boldw_1,\ldots,\Boldw_K$ of $\BoldL_{-l}$ associated to the eigenvalue $0$, with $K\le N-1$ being the number of connected components of $\calG_{-l}$. These eigenvectors have a precise structure: each connected component of $\calG_{-l}$ is associated to an eigenvector with all ones in the entries corresponding to the vertexes of the connected component and all zeros in the remaining entries. 

Since the original graph $\calG$ is connected by assumption, each connected component of $\calG_{-l}$ has at least one vertex adjacent to $l$ in $\calG$. 
Therefore, remembering that $-{\rm diag}({\Bell}_l)$ has ones exactly in the the entires corresponding to the vertexes of $\calG_{-l}$ adjacent to $l$ in $\calG$, this implies 
$-\Boldw_i^T{\rm diag}({\Bell}_l)\Boldw_i>0$ for any $i=1\ldots K$. 

Summarizing, any nonzero vector $\Boldw$ such that $b_1=0$, i.e., $\Boldw\in{\rm ker }\BoldL_{-l}-\{\bf 0\}$ can be expressed as the linear combination $\Boldw=a_1\Boldw_1+\ldots+a_K\Boldw_K$ with at least one $a_i\neq 0$. It then follows that
\[
b_2 = -\Boldw^T{\rm diag}({\Bell}_l)\Boldw= -\sum_{i=1}^Ka_i^2\Boldw_i^T{\rm diag}({\Bell}_l)\Boldw_i>0,
\]
thus concluding the proof of the third item.}

\AFc{Un reviewer dice: I guess that the Laplace expansion formula could be used to simplify the proof of this last item.}
Finally, in order to prove the fourth item, consider any eigenvector $\Boldv$ of $\BoldL_l$ associated to an eigenvalue $\lambda\neq 0$. Since $\BoldL_l$ has a null $l$-th row, the $l$-th component of $\Boldv$ must be necessarily $0$, i.e., $\Boldv=(v_1^T\;0\;v_2^T)^T$. Therefore $\lambda (v_1^T\;0\;v_2^T)^T = \BoldL_l (v_1^T\;0\;v_2^T)^T = 
((\BoldM_{l,1}v_1 + \BoldM_{l,2}v_2)^T \; 0 \; (\BoldM_{l,3}v_1 + \BoldM_{l,4}v_2)^T)^T
$ implying that $\lambda v_1 = 
\BoldM_{l,1}v_1 + \BoldM_{l,2}v_2 $ and $\lambda v_2 = 
\BoldM_{l,3}v_1 + \BoldM_{l,4}v_2$, i.e., $\lambda (v_1^T\;v_2^T)^T = \BoldM_l (v_1^T\;v_2^T)^T$.
\end{proof}

\blue{Since  $\sigma(\BoldL_l) = \sigma(\BoldM_l) \cup \{0\}$, and being $\BoldM_l$ is symmetric,  it follows that  $\BoldL_l$ has \emph{real} eigenvalues, even though it is not symmetric (being $\calG_l$ is a digraph).
Let $0=\lambda_1\le\lambda_2\le\ldots\le\lambda_N$ and $0=\lambda_{1,l}\le\lambda_{2,l}\le\ldots\le\lambda_{N,l}$ be the $N$ real eigenvalues of $\BoldL$ and $\BoldL_l$, respectively.}
 Since $\lambda_2$ is called the `algebraic connectivity' of $\calG$, for similarity we also denote $\lambda_{2,l}$ as the `algebraic connectivity' of the digraph $\calG_l$. From the previous properties we have that, if $\calG$ is connected, then both $\lambda_2>0$ and $\lambda_{2,l}>0$.

\blue{
In order to prove an important property that sheds additional light on the relation between the eigenvalues of $\BoldL$ and $\BoldL_l$ we first  recall a well-known result from linear algebra.
\begin{theorem}[Cauchy Interlace Theorem]\label{thm:cauchy}
Let $\BoldX$ be a Hermitian matrix of order $N$, and let $\BoldY$ be a principal submatrix of $\BoldX$  of order N − 1, i.e., a matrix obtained from  $\BoldX$ by removing any $i$-th row and $i$-th column, with $i\in\{1,\ldots,N\}$. If $\lambda_1^\BoldX \le \lambda_2^\BoldX \le\ldots\le \lambda_{N-1}^\BoldX \le \lambda_N^\BoldX$
lists the eigenvalues of $\BoldX$ and $\lambda_1^\BoldY \le \lambda_2^\BoldY \le\ldots\le \lambda_{N-2}^\BoldY \le \lambda_{N-1}^\BoldY$ the eigenvalues of $\BoldY$, then
$\lambda_1^\BoldX \le \lambda_1^\BoldY \le \lambda_2^\BoldX \le \lambda_2^\BoldY \le\ldots\le \lambda_{N-1}^\BoldX \le \lambda_{N-1}^\BoldY \le \lambda_N^\BoldX$.
\end{theorem}
}
\blue{Then, the following property also holds:}
\begin{property}
For a graph $\calG$ and an induced graph $\calG_l$ it is
$\lambda_{i,l} \le \lambda_i$ for all $i=1\ldots N$.
\end{property}
\begin{proof}
\blue{The property is proven applying Theorem~\ref{thm:cauchy}  the matrixes $\BoldX=\BoldL$ and $\BoldY=\BoldM_l$ and then using the fact that $\sigma(\BoldL_l) = \sigma(\BoldM_l) \cup \{0\}$ thanks to Property~\ref{property:Ml}}.
\end{proof}

To conclude this modeling section we formally prove the stability of the linear system~\eqref{eq:err_p_v_dyn} in the next proposition.
\begin{proposition}\label{prop:stabilty_first_order}
If the graph $\calG$ is  connected, the origin of the linear system~(\ref{eq:err_p_v_dyn}) \blue{with zero input} ($\dot\Boldu_r\equiv 0$) is globally asymptotically stable for any $k_p>0$, $k_u>0$. The rates of convergence of $(\Bolde_p,\,\Bolde_v)$ and $\Bolde_{\hat\Boldu}$ are dictated by $-k_p\lambda_{2,l}$ and $-k_u\lambda_{2,l}$, respectively, where $\lambda_{2,l} = \min \sigma(\BoldM_l)$, i.e., the smallest positive eigenvalue of $\BoldL_l$ (algebraic connectivity of the digraph $\calG_l$). 
\end{proposition}
\begin{proof}
The dynamics of the error $\Bolde$ \blue{with zero input} is:
\begin{align}\footnotesize
\begin{pmatrix}
\dot\Bolde_{\Boldp}\\
\dot\Bolde_{\Boldv}\\
\dot\Bolde_{\hat\Boldu}
\end{pmatrix}
=
\begin{pmatrix}
k_p\BoldG_{l} & \Boldzero_{Nd} & \BoldI_{Nd}\\
\Boldzero_{Nd}  & k_p\BoldG_{l} & k_u\BoldG_{l}\\
\Boldzero_{Nd}  & \Boldzero_{Nd} & k_u\BoldG_{l}
\end{pmatrix}
\begin{pmatrix}
\Bolde_{\Boldp}\\
\Bolde_{\Boldv}\\
\Bolde_{\hat\Boldu}
\end{pmatrix}.
\end{align}
Because of their definition, the sub-vectors $\Bolde_{\Boldp,l}$, $\Bolde_{\Boldv,l}$, and  
$\Bolde_{\hat\Boldu,l}$ (i.e., the errors relative to the agent $l$) are zero at $t=t_0$ and their dynamics is invariant because of the null row in $\BoldL_l$ corresponding to the agent $l$, i.e.,
\[
\Bolde_{\Boldp,l} = \Bolde_{\Boldv,l} =   
\Bolde_{\hat\Boldu,l} = \dot\Bolde_{\Boldp,l} = \dot\Bolde_{\Boldv,l} =   
\dot\Bolde_{\hat\Boldu,l} = \Boldzero, \; \forall t>t_0.
\]
Therefore we can restrict the analysis to the dynamics of the orthogonal subspace, i.e., of the remaining components  $\Bolde_{\Boldp,i}$, $\Bolde_{\Boldv,i}$, and $\Bolde_{\hat\Boldu,i}$ for all $i\neq l$. We denote with ${^l\Bolde}_\Boldp$, ${^l\Bolde}_\Boldv$, and ${^l\Bolde}_{\hat\Boldu}$ the $(N-1)d$-vectors obtained by removing the $d$ entries corresponding to $l$ in $\Bolde_{\Boldp}$, $\Bolde_{\Boldv}$, and $\Bolde_{\hat\Boldu}$, respectively, and with ${^l\Bolde}$ their concatenation. The dynamics of the reduced error $^l\Bolde$ is then: 
\begin{align}\footnotesize
\begin{pmatrix}
{^l\dot\Bolde}_{\Boldp}\\
{^l\dot\Bolde}_{\Boldv}\\
{^l\dot\Bolde}_{\hat\Boldu}
\end{pmatrix}
=
\underbrace{
\begin{pmatrix}
k_p{^l\BoldG_{l}} & \Boldzero_{(N-1)d} & \BoldI_{(N-1)d}\\
\Boldzero_{(N-1)d}  & k_p{^l\BoldG_{l}} & k_u{^l\BoldG_{l}}\\
\Boldzero_{(N-1)d}  & \Boldzero_{(N-1)d} & k_u{^l\BoldG_{l}}
\end{pmatrix}}_{\BoldD_l}
{^l\Bolde},\label{eq:reduced_error}
\end{align}
where $^l\BoldG_{l}=-\BoldM_l\otimes \BoldI_d$. We recall that $\BoldM_l$ is positive definite (see Property~\ref{property:Ml}) and its smallest eigenvalue, denoted as $\lambda_{2,l}$, represents the algebraic connectivity of the digraph associated to $\BoldL_l$. Due to the block diagonal form of $\BoldD_l$ and to the properties of the Kronecker product, the distinct eigenvalues of $\BoldD_l$ are 
at most $2(N-1)$, of which $N-1$ are obtained by multiplying all the eigenvalues of $\BoldM_l$ with $-k_p$ and the remaining $N-1$ by multiplying all the eigenvalues of $\BoldM_l$ with $-k_u$. The thesis then simply follows from the structure of system~(\ref{eq:reduced_error}).
\end{proof}

Therefore, if $\dot{\Boldu}_r\equiv 0$, the agent velocities $\Boldv$ and estimation $\hat\Boldu$ asymptotically converge to the common reference velocity $\Boldu_r$, and the agent positions $\Boldp$ to the desired shape $\Boldone \otimes \Boldp_l  + {\Boldd} - \Boldone \otimes \Boldd_l$. %
Furthermore, the value of $\lambda_{2,l}$ directly affects the convergence rate of the three error vectors $(\Bolde_{\Boldp},\,\Bolde_{\Boldv},\,\Bolde_{\hat\Boldu})$ over time. Since, for a given graph topology $\calG$, $\lambda_{2,l}$ is determined by the \emph{identity} of the leader in the group, it follows that maximization of $\lambda_{2,l}$ over the possible leaders results in a faster convergence of the tracking error. This insight then motivates the {online leader selection} strategy detailed in the rest of the paper.

\section{Effects Of A Changing Leader And Associated Tracking Performance Metric}\label{sec:lead_sel_first}
In this section we provide the second main contribution of this paper by theoretically analyzing how the choice of a changing a leader affects the dynamics of the error vector. \blue{We assume that a new leader can be periodically selected by the group at some frequency $1/T$, $T>0$, and let $t_k=kT$.}
\begin{remark}\label{rem:period_T}
\blue{We note that, in general, the quantities $T$ (the leader election period) and $T_r$ (the reference command period) do not need to be related. However, for the reasons given in Remark~\ref{rem:UUVexample}, it is meaningful to consider $T\leq T_r$ since the internal group communication/dynamics is typically much faster than the master/group interaction. In the following, we then design $T$ to be an exact divisor of $T_r$, i.e., such that $T_r/T\in\mathbb{N}$.}
\end{remark}
Let us also denote the leader at time $t_k$ with the index $l_k$, \blue{and recall that the velocity reference $\Boldu_r$, between $t_k$ and $t_{k+1}$ is constant (see Remark~\ref{rem:UUVexample}).}
Rewriting the dynamics of system~(\ref{eq:est_ref_vel})--(\ref{eq:EgerFixed})  among consecutive \blue{leader-selection} times, i.e.,  during the interval $[t_k,t_{k+1})$, we obtain:
\begin{align}
\dot{\hat\Boldu} &= k_u\BoldG_{l_k}\hat\Boldu & t\in[t_k,t_{k+1})\label{eq:dynamics_u}\\
\dot{\Boldp} &= \hat\Boldu +k_p\BoldG_{l_k}(\Boldp - \Boldd) & t\in[t_k,t_{k+1})\label{eq:dynamics_p}
\end{align}
with initial conditions
\begin{align}
\hat\Boldu(t_k) &= \hat\Boldu(t_k^-) + (\bar \BoldS_{l_k}\otimes I_d)(\Boldone \otimes \Boldu_r(t_k) -\hat\Boldu(t_k^-))\label{eq:init_cond_u}\\
\Boldp(t_k) &= \Boldp(t_k^-),\label{eq:init_cond_p}
\end{align}
and, for the velocity vector $\Boldv$,
\begin{align}
\Boldv(t_k) &= \hat\Boldu(t_k) +k_p\BoldG_{l_k} (\Boldp(t_k) - \Boldd). \label{eq:init_cond_v} 
\end{align}
Matrix $\bar \BoldS_{l_k}\in\mathbb{R}^{N\times N}$ is a diagonal selection matrix with all zeros on the main diagonal but the $l_k$-th entry set to one, and its complement is defined as $\BoldS_{l_k}=\BoldI_N - \bar \BoldS_{l_k}$.

Equation~\eqref{eq:init_cond_u} represents the reset action~(\ref{eq:vel_estimator_reset}) performed on the components of $\hat\Boldu$ corresponding to the new leader $l_k$ which are reset to $\Boldu_r(t_k)$. The initial condition $\hat\Boldu(t_k)$ hence depends on the chosen leader $l_k$ and is in general discontinuous at $t_k$. Similar considerations hold for the value of the velocity vector $\Boldv(t_k)$. %
On the other hand, the position vector $\Boldp(t)$ is continuous at $t_k$.  %

Focusing on the error dynamics~(\ref{eq:err_p_v_dyn}) during the interval $[t_k,t_{k+1})$, and noting that $\Boldu_r(t)\equiv {\rm const}$ in this interval by assumption, we obtain
\begin{align}
\dot\Bolde\footnotesize
=
\begin{pmatrix}
k_p\BoldG_{l_k} & \Boldzero_{Nd} & \BoldI_{Nd}\\
\Boldzero_{Nd}  & k_p\BoldG_{l_k} & k_u\BoldG_{l_k}\\
\Boldzero_{Nd}  & \Boldzero_{Nd} & k_u\BoldG_{l_k}
\end{pmatrix}
\Bolde.
\end{align}
Using~(\ref{eq:init_cond_u}--\ref{eq:init_cond_v}), the initial conditions at $t_k$ for $\Bolde = (\Bolde_{\Boldp}^T \; \Bolde_{\Boldv}^T \; \Bolde_{\hat\Boldu}^T)^T$ as a function of the chosen leader $l_k$ and of the received external command $\Boldu_r(t_k)$ are then: %
\begin{align}
\Bolde_\Boldp(t_k) &=%
(\BoldS_{l_k}\otimes I_d)(\Boldp(t_k^-)-\Boldd - \Boldone\otimes (\Boldp_{l_{k}}(t_k^-)- \Boldd_{l_{k}})
\label{eq:initial_error_p}\\
\Bolde_\Boldv(t_k) &=%
(\BoldS_{l_k}\otimes I_d)(\hat\Boldu(t_k^-) - \Boldone\otimes \Boldu_r(t_k) + \Boldgamma(t_k^-))\label{eq:initial_error_v}\\
\Bolde_{\hat\Boldu}(t_k) &=%
(\BoldS_{l_k}\otimes I_d)(\hat\Boldu(t_k^-) - \Boldone\otimes \Boldu_r(t_k))\label{eq:initial_error_u}
\end{align}
where
$\Boldgamma=-k_p(\BoldL\otimes I_d)(\Boldp - \Boldd)$, i.e., $\Boldgamma=\begin{pmatrix}\Boldgamma_1^T\ldots\Boldgamma_N^T\end{pmatrix}^T\in\mathbb{R}^{Nd}$ and
\[
\Boldgamma_i=k_p\sum_{j\in\calN_i}\left((\Boldp_j-\Boldp_i) - (\Boldd_j-\Boldd_i)\right).
\] 
Therefore, from~(\ref{eq:initial_error_p}--\ref{eq:initial_error_u}) it follows that vector $\Bolde(t_k)$ is directly affected by the choice of $l_k$. %
For this reason, whenever appropriate we will use the notation $\Bolde(t_k,\,l_k)$ to explicitly indicate this (important) dependency.
We also note that $\Boldgamma$ depends on $\BoldL$ and not on $\BoldL_{l_k}$.

The following lemma is preliminary to the main result of the section.

\begin{lemma}\label{lem:Q_eigen}
Consider any symmetric matrix $\BoldF\in\mathbb{R}^{M\times M}$ and three positive gains $k_n,k_p,k_u>0$. Denote with $\lambda_1\ldots\lambda_M\in\mathbb{R}$ the  eigenvalues of $\BoldF$. Then define the symmetric matrix
\[
\BoldQ=\footnotesize\begin{pmatrix}
\frac{k_p}{k_n}\BoldF & \Boldzero_{M} & \frac{1}{2k_n}\BoldI_{M}\\
\Boldzero_{M}  & \frac{k_p}{k_n}\BoldF & \frac{k_u}{2k_n}\BoldF\\
 \frac{1}{2k_n}\BoldI_{M}  & \frac{k_u}{2k_n}\BoldF & k_u\BoldF
\end{pmatrix}.
\] The following facts hold:
\begin{enumerate}
\item the $3M$ eigenvalues of $\BoldQ$
are
\begin{align}%
\mu_{1}(\lambda_i) &= \frac{k_p}{k_n}\lambda_i \label{Q_eigen_1}\\ 
\mu_{2}(\lambda_i) &=
\frac{
\lambda_ik_1
-
\sqrt{
1+\lambda_i^2 k_2
}
}{2k_n} \label{Q_eigen_2}\\
\mu_{3}(\lambda_i) &=
\frac{
\lambda_ik_1
+
\sqrt{
1+\lambda_i^2 k_2
}
}{2k_n}\label{Q_eigen_3}
\end{align}
for all $i=1\ldots M$, with $k_1=k_p+k_nk_u$ $(>0)$ and $k_2 = k_u^2+ (k_p-k_nk_u)^2$ $(>0)$;
\item if $\lambda_1\le\lambda_2\ldots\le\lambda_M <0$ and $k_n$ is chosen such that 
\begin{align}
\lambda_M^2(4k_nk_pk_u - k_u^2)>1\label{eq:Q_condition}
\end{align}
then $\mu_j(\lambda_i)<0$ for all $j=1,2,3$, $i=1\ldots M$, and
\[
\mu_3(\lambda_M) = \max_{\substack{j=1,2,3\\i=1\ldots M}}\mu_j(\lambda_i)
\]
\end{enumerate}
\end{lemma}
\begin{proof}
We first prove item 1). 	For any eigenvalue $\mu$ of $\BoldQ$ it holds 
\begin{align}
\BoldQ v = \mu v\label{eq:Q_eigen}
\end{align} 
where $v=(v_1^T\;v_2^T\;v_3^T)^T\in\mathbb{R}^{3M}$ is a unit-norm eigenvector of $\BoldQ$ associated to $\mu$. 
Consider the matrix $x^T\otimes \BoldI_3
\in\mathbb{R}^{3\times 3N}$, where $x_i\in\mathbb{R}^M$ is a unit-norm eigenvector of $\BoldF$ associated to any eigenvalue $\lambda_i$ of $\BoldF$, $i=1\ldots M$.
Left-multiplying both sides of~\eqref{eq:Q_eigen} with $x_i^T\otimes \BoldI_3$ and exploiting the symmetry of $\BoldF$, we obtain 
\begin{align}\tiny
(x_i^T\otimes \BoldI_3) \BoldQ v = 
\underbrace{\begin{pmatrix}
\frac{k_p}{k_n}\lambda_i & 0 & \frac{1}{2k_n}\\
0  & \frac{k_p}{k_n}\lambda_i & \frac{k_u}{2k_n}\lambda_i\\
\frac{1}{2k_n}  & \frac{k_u}{2k_n}\lambda_i & k_u\lambda_i
\end{pmatrix}}_\text{$\BoldQ_{\lambda_i}$}
\begin{pmatrix}
x_i^Tv_1\\
x_i^Tv_2\\
x_i^Tv_3
\end{pmatrix} 
=
\mu \begin{pmatrix}
x_i^Tv_1\\
x_i^Tv_2\\
x_i^Tv_3
\end{pmatrix}.\notag
\end{align}
Therefore $\mu$ is also an eigenvalue of the 3-by-3 matrix $\BoldQ_{\lambda_i}$ for every $\lambda_i\in\sigma(\BoldF)$ $i=1\ldots M$. 
In particular, after some straightforward algebra, this implies that all the eigenvalues of $\BoldQ$ are the solutions of $M$ cubic equations of the form:
\[
\left(
\mu -
\tfrac{k_p\lambda_i}{k_n}
\right)
\left(
\mu^2 
-\tfrac{\lambda_i(k_p +k_nk_u)}{k_n}\mu
-\tfrac{\lambda_i^2(k_u^2 - 4k_nk_pk_u) +1}{4k_n^2}
\right)=0,
\]
for $\lambda_i=1\ldots M$, which then leads to~(\ref{Q_eigen_1}-\ref{Q_eigen_3}) and proves item 1).

We now prove the item 2). 

First of all, under the stated conditions, it is $\mu_3(\lambda_i)>\mu_j(\lambda_i)$ for any $j=1,2$ and $i=1\ldots M$, and $\mu_3(\lambda_i)>\mu_2(\lambda_i)$ follows from $\lambda_i<0$ and $k_1,k_n>0$.
On the other hand, the inequality $\mu_3(\lambda_i)>\mu_1(\lambda_i)$ can be shown, after some algebra, being equivalent to
\[
\sqrt{1+\lambda_i^2k_u^2 + \lambda_i^2(k_p-k_nk_u)^2}>\lambda_i(k_p-k_nk_u),
\]
which holds for any value of $\lambda_i$. 
Therefore the negativity of the eigenvalues of $\BoldQ$ is guaranteed by the negativity of $\mu_3(\lambda_i)$, for every $i=1\ldots M$. Condition $\mu_3(\lambda_i)<0$, after straightforward algebra, is equivalent to $\lambda_i^2(4k_nk_pk_u - k_u^2)>1$, for every $i=1\ldots M$. Furthermore, since $\lambda_M$ has the smallest absolute value among the eigenvalues of $\BoldF$, it is sufficient to guarantee that $\lambda_M^2(4k_nk_pk_u - k_u^2)>1$, which proves the first part of fact 2).

In order to prove the second part, it is sufficient to show that $\mu_3(\lambda_M) > \mu_3(\lambda_i)$ for any $i\neq M$. To this end, we prove that $\mu_3(\lambda_i)$ is a monotonically increasing function of $\lambda_i$ in the interval $(-\infty,-\frac{1}{\sqrt{4k_nk_pk_u - k_u^2}})$, and has therefore its maximum for $i=M$. By simple derivation we obtain 
\begin{align}
\frac{\partial\mu_3}{\partial \lambda_i} = 
\frac{1}{2k_n}
\left(
k_1 + \frac{k_2 \lambda_i}{\sqrt{1+k_2\lambda_i^2}}
\right)\notag
\end{align}
which can be positive (after some algebra) if and only if 
\begin{align}
k_1^2 + k_2(k_1^2-k_2)\lambda_i^2>0.\label{eq:pos_deriv}
\end{align}
Noting that $k_1^2-k_2 = 4k_pk_uk_n-k_u^2$ and applying~\eqref{eq:Q_condition} we obtain $(k_1^2-k_2)\lambda_i^2>1$, which implies that~\eqref{eq:pos_deriv} is always satisfied under our assumptions, then concluding the proof of item 2).
\end{proof}

The following result gives an explicit characterization of the behavior of  $\Bolde(t)$ during the interval $[t_k,t_{k+1})$.
\begin{proposition}\label{prop:error_bound}
Consider the error metric \AFc{un reviewer suggerisce di usare $\Vert {\Bolde} \Vert^2_{P_{k_n}} $ TOO LONG?}
\begin{equation}\label{eq:err_metric_first_order}
\Vert {\Bolde} \Vert^2_{k_n} 
=%
{\Bolde}^T
\underbrace{
\begin{pmatrix}
\BoldI_{Nd}/k_n & \Boldzero_{Nd} & \Boldzero_{Nd}\\
\Boldzero_{Nd} & \BoldI_{Nd}/k_n & \Boldzero_{Nd}\\
\Boldzero_{Nd} & \Boldzero_{Nd} & \BoldI_{Nd}
\end{pmatrix}}_{=:\BoldP_{k_n}}
{\Bolde},
\end{equation}
with $k_n>0$. For any pair of positive gains $k_p$ and $k_u$, if $k_n$ is chosen such that $\lambda_{2,l}^2(4k_nk_pk_u - k_u^2)>1$ then, in closed-loop, 
$\Vert {\Bolde}(t) \Vert^2_{k_n}$ monotonically decreases during the interval $[t_k,t_{k+1})$, being in particular dominated by the exponential upper bound:
\begin{align}
\Vert {\Bolde}(t) \Vert^2_{k_n} \leq \Vert \Bolde (t_k) \Vert^2_{k_n} \, e^{-2\;{\mu}_{l_k}(t-t_k)} \quad \forall t\in [t_k,t_{k+1}),\label{eq:square_err1}
\end{align}
where
\begin{align}
\mu_{l_k} =
\frac{
{\lambda}_{2,l_k}k_1
-
\sqrt{
1+{\lambda}_{2,l_k}^2k_2
}
}{2k_n}>0
\label{eq:mu_l}
\end{align}
with $k_1=k_p+k_nk_u$ and $k_2=k_u^2 + (k_p-k_nk_u)^2$.
\end{proposition}
\begin{proof}
Adopting the same arguments of the proof of Prop.~\ref{prop:stabilty_first_order} during the interval $[t_k,t_{k+1})$, and omitting (as in the following) the dependency upon the time-step $k$, we obtain a dynamics of the reduced error $^l\Bolde$, in the interval $[t_k,t_{k+1})$ equivalent to~\eqref{eq:reduced_error}.

Notice that clearly
\[
\|\Bolde\|^2_{k_n} = {\Bolde}^T\BoldP_{k_n}{\Bolde} = {^l\Bolde^T}{^l\BoldP_{k_n}}{^l\Bolde} = \|^l\Bolde\|^2_{k_n},
\] where ${^l\BoldP_{k_n}}$ is a $d(N-1)\times d(N-1)$ matrix obtained by removing the $d$ columns and rows of ${\BoldP_{k_n}}$ corresponding to $l$. 

Consider now the dynamics of $\|\Bolde\|^2_{k_n}=\|^l\Bolde\|^2_{k_n}$:
\begin{align}
\begin{aligned}
\frac{d}{dt}\|^l\Bolde\|^2_{k_n} &= 2{^l\Bolde^T}{^l\BoldP_{k_n}}{^l\dot\Bolde} = 2{^l\Bolde^T}{^l\BoldP_{k_n}}\BoldD_l {^l\Bolde}=\\
& = 
2{^l\Bolde^T}{\rm sym}({^l\BoldP_{k_n}}\BoldD_l){^l\Bolde} \le 2\mu_{{\rm max},l} \|^l\Bolde\|^2, 
\end{aligned}\label{eq:norm_dynamics}
\end{align}
with $\mu_{{\rm max},l}$ being the largest eigenvalue of the symmetric part of ${^l\BoldP_{k_n}}\BoldD_l$, i.e., of
\[
{\rm sym}({^l\BoldP_{k_n}}\BoldD_l) 
=\footnotesize
\begin{pmatrix}
\frac{k_p}{k_n}{^l\BoldG_{l}} & \Boldzero_{(N-1)d} & \frac{\BoldI_{(N-1)d}}{2k_n}\\
\Boldzero_{(N-1)d}  & \frac{k_p}{k_n}{^l\BoldG_{l}} & \frac{k_u{^l\BoldG_{l}}}{2k_n}\\
 \frac{\BoldI_{(N-1)d}}{2k_n}  & \frac{k_u{^l\BoldG_{l}}}{2k_n} & k_u{^l\BoldG_{l}}
\end{pmatrix}.
\]
Equation~\eqref{eq:norm_dynamics} implies that $\forall\;t\in [t_k,t_{k+1})$
\begin{align}
\Vert {\Bolde}(t) \Vert^2_{k_n} \leq \Vert \Bolde (t_k) \Vert^2_{k_n} \, e^{2\;{\mu}_{{\rm max},l_k}(t-t_k)}\label{eq:square_err10}.
\end{align}
We then show that ${\mu}_{{\rm max},l}=-{\mu}_{l}$, where ${\mu}_{l}$ is given in~\eqref{eq:mu_l}.

First of all note that, due to the properties of the Kronecker product,  the eigenstructure of ${\rm sym}({^l\BoldP_{k_n}}\BoldD_l) $ is obtained by repeating $d$ times the one of
\[
{^l\BoldQ_l} = \footnotesize
\begin{pmatrix}
-\frac{k_p}{k_n}{^l\BoldM_{l}} & \Boldzero_{N-1} & \frac{\BoldI_{(N-1)}}{2k_n}\\
\Boldzero_{N-1}  & -\frac{k_p}{k_n}{^l\BoldM_{l}} & -\frac{k_u{^l\BoldM_{l}}}{2k_n}\\
 \frac{\BoldI_{(N-1)}}{2k_n}  & -\frac{k_u{^l\BoldM_{l}}}{2k_n} & -k_u{^l\BoldM_{l}}
\end{pmatrix}.
\]
Applying Lemma~\ref{lem:Q_eigen} with $\BoldA=-\BoldM_l$ and thus  $\lambda_M=-\lambda_{2,l}$, it follows that, if $k_n$ is chosen such that $\lambda_{2,l}^2(4k_nk_pk_u - k_u^2)>1$, then $-\mu_{l}=\mu_3(-\lambda_{2,l})=\mu_{{\rm max},l}$ is the largest eigenvalue of $^l\BoldQ_l$, thus finally proving the proposition.
\end{proof}

{Note that Prop.~\ref{prop:error_bound} proves that the scalar metric $\Vert {\Bolde} \Vert^2_{k_n} $ is monotonically decreasing along the system trajectories, while this may not hold for other metrics such as the canonical $\Vert {\Bolde} \Vert^2$}.
\blue{Since $\Vert {\boldsymbol e} \Vert^2_{k_n}$ is monotonically decreasing along the system trajectories, regardless of the current leader, it also constitutes a common Lyapunov function for the switching system~\cite{2003-Lib}. Therefore the stability of the system under changing leaders is also guaranteed.}

Furthermore, Prop.~\ref{prop:error_bound} provides a very important results since, at every $t=t_k$, the bound~(\ref{eq:square_err1}) allows to compute an  estimation of the future decrease of the error vector $\Bolde(t)$ in the interval $[t_k,t_{k+1})$. In particular, by evaluating~(\ref{eq:square_err1}) at $t=t^-_{k+1}$, i.e., just before the next leader selection, we obtain
\begin{align}
\Vert \Bolde(t^-_{k+1}) \Vert^2_{k_n} &\leq \Vert \Bolde(t_k,\,l_k) \Vert^2_{k_n} \, e^{-2\;\mu_{l_k} T}. 
 \label{eq:cost1}
\end{align}
Since both $\Bolde(t_{k},\,l_k)$ and $\mu_{l_k}$ depend on the value of $l_k$ (i.e., the identity of the leader), the rhs of~\eqref{eq:cost1} can be exploited for choosing the leader at time $t_k$ in order to maximize the convergence rate of $\Bolde(t)$ during the interval $[t_k,t_{k+1})$ and therefore improving, at the same time, both the tracking of the reference velocity $\Boldu_r(t)$ and of the desired formation encoded by $\Boldd$. %

These remarks are formalized by the following statement.
\begin{corollary}%
\label{prob:lead_sel_first}
In order to improve the tracking performance of the reference velocity and of the desired formation during the interval $t\in[t_k,\,t_{k+1})$, the group should select the leader that solves the following minimization problem %
\begin{align}
\arg\min_{l\in\calL_k}\Vert {\Bolde}(t_k,\,l) \Vert^2_{k_n} \, e^{-2\;\mu_l T},\label{eq:minimization_first}
\end{align}
{where $\calL_k$ is the set of `eligible' agents from which a leader can be selected at $t_k$}.
\end{corollary}

\AFc{Spiegare meglio perché c'è l'esponenziale Un reviewer non ha capito perché dice: \emph{Doesn't it make sense to simply select
the node taking the largest error (without the exponential part)
at time $t_k$ so that that agent will have better control for the
next time period?}}

\AFc{Ridefinire meglio il problema in modo che sia chiaro cosa vogliamo risolvere (2 reviewers lo dicono). Un reviewer dice: \emph{While optimal leader selection problem is formulated, it is not explained how each agent se- lects leader by solving the optimization problem. After introducing leader selection problem in Definition 1, it is claimed that the agents can maximize online the tracking performance. However, there is no explanation how this is done for each agent. In view of the fact that it is a nonconvex optimization problem involving integer constraints, the authors should provide a detailed optimization algorithm for the leader selection problem.}}

\begin{remark}
Note that, in the cost function~(\ref{eq:minimization_first}), both $\Vert {\Bolde}(t_k,\,l) \Vert^2_{k_n}$ and $e^{-2\;\mu_l T}$ depend on the chosen leader $l$. 
{Therefore the minimization problem~(\ref{eq:minimization_first}) can only be solved \emph{online} since the cost function depends on both the group topology and the current agent state.}
\end{remark}

\begin{remark}
We note that, because of the reset actions performed in~(\ref{eq:vel_estimator_reset}) and~\eqref{eq:EgerFixed}, every instance of the leader selection potentially leads to a decrease of $\Vert \Bolde(t_k,\,l_k) \Vert^2_{k_n}$ since it zeroes the $l_k$ $d$-components of the estimation and velocity error vectors $\Bolde_{\hat u}$, $\Bolde_v$. Therefore, it would be desirable to reduce as much as possible the selection period $T$. In practice, however, there will exist a finite minimum selection period $T\geq T_{min}>0$ upper bounding the highest frequency at which the leader selection process can be reliably executed (because of, e.g., the limited bandwidth capabilities of the multi-agent group). %
\end{remark}

\section{Decentralized Computation Of The Next Best Neighboring Leader}\label{sec:decentr_algo}

{In order to obtain a global optimum,~\eqref{eq:minimization_first} should be minimized among all the agents in the group, i.e., by setting $\calL_k=\{1,\ldots,N\}$. However this would result in a fully centralized optimization problem. \green{Since we aim for a \emph{decentralized} solution, in this section we consider a decentralized (sub-optimal) version where~\eqref{eq:minimization_first} is solved only among the $1$-hop neighbors of the current leader $l_k$, i.e., by setting $\calL_k = \calN_{l_k}$.}
Nevertheless, even in this `decentralized' case, evaluating~\eqref{eq:minimization_first} for each $l\in\calL_k$ requires to compute two global quantities  for each $l$, i.e., $\Vert {\Bolde}(t_k,\,l) \Vert^2_{k_n}$ and $\mu_l$. We then now provide the third main contribution of this paper by showing how to render this computation fully distributed, i.e., only relying on local and $1$-hop information available to the master and to the current leader.

Let us then consider the evaluation of $\Vert {\Bolde}(t_k,\,m) \Vert^2_{k_n} \, e^{-2\;\mu_m T}$ in~\eqref{eq:minimization_first} by a candidate agent $m\in\calL_{k-1}$. 
This requires knowledge of two global quantities: the error norm $\Vert {\Bolde}(t_k,\,m) \Vert^2_{k_n}$ and the connectivity eigenvalue $\lambda_{2,m}$ of digraph $\calG_m$ for computing  $\mu_m$ via~\eqref{eq:mu_l}. An estimation of the value of $\lambda_{2,m}$ can be obtained in a decentralized way by employing a simplified version of the \emph{Decentralized Power Iteration algorithm} proposed in~\cite{2010-YanFreGorLynSriSuk} without the deflation step (since $\lambda_{2,m}$ is the smallest eigenvalue of the matrix $\BoldM_m$, which in fact does not possess a structural eigenvalue in zero as it is for $\BoldL$).
It is well known that a possible issue of the power iteration is the speed of convergence for large networks. For static network this does not represent a problem since the distributed power iteration can be run just once at the beginning before starting the task. The method can be still applied for a slowly time-varying network if the parameters (e.g., the gains) of the distributed power iteration are tuned in advance depending on the variability and the speed of the network (see, e.g.,~\cite{2013e-RobFraSecBue} for a use of the distributed power iteration in the case of time-varying graphs). 

\AFc{Qui un reviewer dice che la power iteration si sa che è lenta, e che quindi non si addice al nostro problema che deve essere risolto online con grafo tempo-variante. e poi fa queste domande:\\
- How fast and how accurate can each agent estimate the performance metric?\\
- What happens if different agents have different estimates of the performance metric?Would the estimation error results in a contradiction between different agents as what
leader should be chosen?\\
- How to guarantee that all agents agree on the solution to the optimization problem?}

\AFc{Un altro reviewer dice: \emph{The distributed algorithm is required to compute $\lambda_2$ but this
must be expensive if the network is large. I wonder if the time
delay for making this computation is considered in the simulation,
the performance will be the same.}}

\begin{proposition}
The scalar quantities $\Vert \Bolde(t_k,\,m) \Vert^2_{k_n}$ for $m\in\calL_{l_{t_{k-1}}}$ can be evaluated by the previous leader $l_{k-1}$ in a decentralized way by resorting to local computation and distributed estimation.
\end{proposition}
\begin{proof}
We first note that the quantities $\Boldp_m$, $\hat\Boldu_m$ and $\Boldgamma_m$ are locally available to agent $m$, while $\Boldu_r$ can be retrieved from the current leader $l_{k-1}$ via $1$-hop communication. It is then convenient to expand $\|\Bolde(t_k,\,m)\|^2_{k_n}$ as:
\begin{equation}\label{eq:err_dec}
\|\Bolde\|^2_{k_n}= \Bolde^T_{\hat\Boldu}\Bolde_{\hat\Boldu}+\frac{1}{k_n}\Bolde^T_\Boldp\Bolde_\Boldp+\frac{1}{k_n}\Bolde^T_\Boldv\Bolde_\Boldv,
\end{equation}
where we omitted the various dependencies for brevity. For every vector $(S_m\otimes I_d)\Boldx$, it is 
$$
\|(S_m\otimes I_d)\Boldx\|^2= \sum_{i=1}^N\|\Boldx_i\|^2 - \|\Boldx_{m}\|^2.
$$
Denoting with the  superscript~$^-$ the quantities computed at $t_k^-$, and using~(\ref{eq:initial_error_p}--\ref{eq:initial_error_u}), the three terms in~\eqref{eq:err_dec} can then be rewritten as
\begin{align}
\begin{aligned}
&\Bolde^T_{\hat\Boldu}\Bolde_{\hat\Boldu}=\sum_{i=1}^N
\|\hat\Boldu_i^- - \Boldu_r\|^2 - \|\hat\Boldu_m^- - \Boldu_r\|^2 = \notag\\
& \sum_{i=1}^N
\hat\Boldu_i^{-T}\hat\Boldu_i^- - 2\Boldu_r^T\sum_{i=1}^N\hat\Boldu_i^- + N\Boldu_r^T\Boldu_r  - \|\hat\Boldu_m^- - \Boldu_r\|^2
\end{aligned}
\end{align}
and
\begin{align}%
\begin{aligned}
&\Bolde^T_\Boldv\Bolde_\Boldv=\sum_{i=1}^N
\|\hat\Boldu_i^- - \Boldu_r +\Boldgamma_i^-\|^2 - \|\hat\Boldu_m^- - \Boldu_r+\Boldgamma_m^-\|^2 = \\
& \sum_{i=1}^N
\hat\Boldu_i^{-T}\hat\Boldu_i^- + \sum_{i=1}^N
\Boldgamma_i^{-T}\Boldgamma_i^- + N\Boldu_r^T\Boldu_r - 2\Boldu_r^T\sum_{i=1}^N\hat\Boldu_i^-+\\
& 2\sum_{i=1}^N\Boldu_i^{-T}\Boldgamma_i^-  - 2\Boldu_r^T\sum_{i=1}^N\Boldgamma_i^- - \|\hat\Boldu_m^- - \Boldu_r + \Boldgamma_m^-\|^2.
\end{aligned}\label{eq:e_v_norm_distr}
\end{align}
We can further simplify~\eqref{eq:e_v_norm_distr} by noting that, being $\Boldone^TL=\Boldzero$, it is  $- 2\Boldu_r^T\sum_{i=1}^N\Boldgamma_i^- = \Boldzero$. Finally, letting $\tilde\Boldp^- 
= \Boldp^- - \Boldd$, we obtain 
\begin{align}
\Bolde^T_\Boldp\Bolde_\Boldp &= \sum_{i=1}^N
\|\tilde\Boldp_i^- - \tilde\Boldp_m^-\|^2 + 0 =\notag\\
&= \sum_{i=1}^N
\tilde\Boldp_i^{-T}\tilde\Boldp_i^-
- 2\tilde\Boldp_m^{-T}\sum_{i=1}^N\tilde\Boldp_i^- 
  + N\tilde\Boldp_m^{-T}\tilde\Boldp_m^-. \notag
\end{align}

Therefore, we can conclude that the quantity $\Vert \Bolde(t_k,\,m) \Vert^2_{k_n}$ can be evaluated by agent $m$ as a function of:
\begin{enumerate}
\item the vectors $\Boldp_m(t_k^-)$, $\hat\Boldu_m(t_k^-)$ and $\Boldgamma_m(t_k^-)$ (locally available to agent $m$);
\item the vector $\Boldu_r(t_k)$ (available to $m$ via communication from the current leader $l_{k-1}$);
\item the three vectors $\sum_{i=1}^N\hat\Boldu_i(t_k^-)$, $\sum_{i=1}^N\Boldp_i(t_k^-)$, and $\sum_{i=1}^N(\Boldp_i(t_k^-)-\Boldd_i)$ (not locally available to agent $m$), 
\item the four scalar quantities $\sum_{i=1}^N \hat\Boldu_i^{-T}\hat\Boldu_i^-$, $\sum_{i=1}^N\Boldgamma_i^{-T}\Boldgamma_i^-$, $\sum_{i=1}^N\Boldu_i^{-T}\Boldgamma_i^-$, and $\sum_{i=1}^N\tilde\Boldp_i^{-T}\tilde\Boldp_i^-$ (not locally available to agent $m$),
\item the total number of agents $N$.
\end{enumerate}

The three vectors and four scalar quantities listed in 3)--4) cannot be retrieved using only local and $1$-hop information. However, a decentralized estimation of their values can be obtained by resorting to the \emph{PI-ace} filtering technique introduced in~\cite{2006-FreYanLyn}. In fact, given a generic vector quantity $\Boldx\in\calbR^N$ with every component $\Boldx_i$ locally available to agent $i$, the PI-ace filter allows every agent in the group to build an estimation converging to the average $\sum^N_{i=1}\Boldx_i/N$.
 
\AFc{qui un revisore dice che dobbiamo spiegare meglio bene come queste quantità si possono ottenere con il PI-ace o in generale in modo decentralizzato, queste sono le sue domande:\\
- What is the basic idea behind this PI-ace technique?\\
- How to do it in decentralized manner?\\
- How to guarantee consensus over agents of these desired quantities?\\
- How the estimation error (resulted from PI-ace technique) affects the selection of leader?}

If $N$ is known, the total sum $\sum^N_{i=1}\Boldx_i$ can then be immediately recovered, otherwise %
it is nevertheless possible to resort to an additional decentralized scheme (see, e.g.,~\cite{2011-BriZelBurAll}) to obtain its value over time.
Therefore, this analysis allows to conclude that agent $m$ can estimate the various quantities listed in points 3)--4), and thus compute $\Vert \Bolde(t_k,\,m) \Vert^2_{k_n}$, in a decentralized way. 
\end{proof}

For the reader convenience we summarize in Algorithm~\ref{algo:1} the decentralized ``Online Leader Selection'' run by the agents at every $t_k$, where $\hat{c}_m[k]=\Vert {\Bolde}(t_k,\,m) \Vert^2_{k_n} \, e^{-2\;\mu_m T}$ denotes the cost function in~(\ref{eq:minimization_first}) evaluated for $l=m$

\begin{algorithm}[t]
\caption{{\it Decentralized Online Leader Selection}}

\nllabel{algo:1}                  

\blue{

\small{

Denote with $l_0$ the first selected leader (e.g., randomly)\;

$k\leftarrow 1$\;

\While{true}
{

  \If{$(k-1)T/T_r \in \mathbb{N}_0$ }
  {
  
    agent $l_{k-1}$ informs the master about its leadership\;
  
    agent $l_{k-1}$ receives a new value of $\Boldu_r(t_k)$ from the master\;

    agent $l_{k-1}$ sends $\Boldu_r(t_k)$ to every neighbor in $\calN_{k-1}$\; 
  }
	
  every agent $m\in \calN_{l_{k-1}}$ sends $\hat c_m[k]$ to agent $l_{k-1}$\;

  agent $l_{k-1}$ computes the set $C_{k} = {\rm argmin}_{m\in\calL_{k-1}} \hat c_m[k]$\;

	\eIf{$l_{k-1} \in C_{k+1}$}
	{
		$l_{k}=l_{k-1}$
	}{
		agent $l_{k-1}$ nominates $l_{k}$ in $C_{k}$, e.g., randomly\;
	}

	keep implementing the distributed controllers and estimators until $T$ elapses\;

	$k\leftarrow k+1$\;
}

}

}

\end{algorithm}

\begin{figure}
\centering
\includegraphics[width=0.35\textwidth]{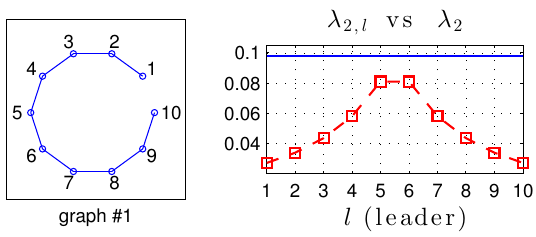}
\includegraphics[width=0.35\textwidth]{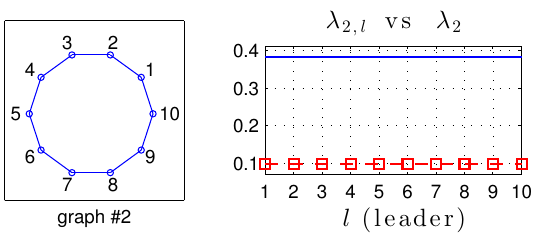}
\includegraphics[width=0.35\textwidth]{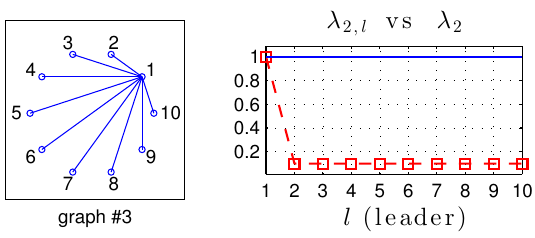}
\includegraphics[width=0.35\textwidth]{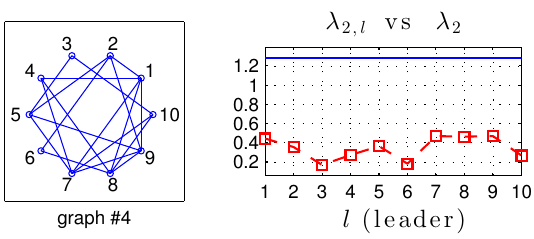}
\includegraphics[width=0.35\textwidth]{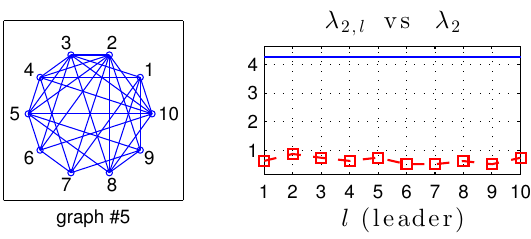}
\includegraphics[width=0.35\textwidth]{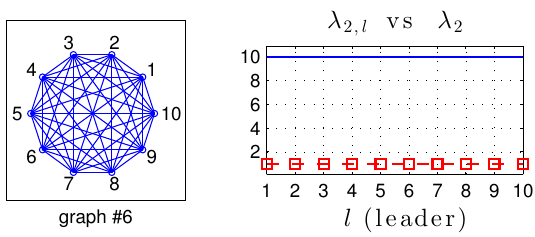}
\caption{Values of $\lambda_{2,l}$ vs. $\lambda_2$ for different leaders $l$. The squares correspond to values of $\lambda_{2,l}$ associated to a leader $l=1\ldots N$, with $N=10$. The solid constant blue lines represent $\lambda_2$.
Each row corresponds to a different graph with $N=10$ vertexes. From the top to the bottom: the line, ring, star, two random (connected) graphs, and a complete graph.}
\label{fig:canonical_graphs}
\end{figure}

\section{Numerical Examples}\label{sec:simul}

We report now some numerical results 
meant to illustrate the effectiveness of the proposed approach.
We compare four different leader selection strategies: $(i)$ no leader selection (thus, constant leader during task execution); $(ii)$ the decentralized leader selection summarized by Algorithm~\ref{algo:1}; $(iii)$ a \emph{globally informed} variant of Algorithm~\ref{algo:1} where, at each iteration, the leader is selected as the one minimizing~\eqref{eq:minimization_first} among \emph{all} the agents in the group rather than within the set $\calL_{k-1}$ of leader neighbors; $(iv)$ a \emph{random} leader selection. All the four runs started from the same initial conditions and involved a group of $N=10$ agents. The interaction graph $\calG$ was cycled over the six topologies shown in Fig.~\ref{fig:canonical_graphs} with a switching frequency of $2$~s, and the velocity command $\Boldu_r$ was received by the current leader with a sending period $T_r=5$~s. Finally, the leader selection algorithm was executed with period $T=0.05$~sec, and the gains $k_p=5$, $k_u=2.5$ were employed. Note that the algorithm result does not depend on the particular shape defined by $\Boldd$ therefore we just selected an arbitrary $\Boldd$ for the examples.

Figures~\ref{fig:simulation_first_order}(a--e) report the results of the four simulation runs: Fig.~\ref{fig:simulation_first_order}(a)-top shows the current graph $\calG$ topology during the simulations (according to the indexing used in Fig.~\ref{fig:canonical_graphs}) and Fig.~\ref{fig:simulation_first_order}(a)-bottom the behavior of $\Boldu_r(t)$ which, as expected, is piece-wise constant and has a jump at every $T_r$~sec.  The four Figs.~\ref{fig:simulation_first_order}(b--e) then report the behavior of $l(t)$ (the identity of the current leader) and of $\|\Bolde(t)\|_{k_n}$, the error metric defined in~(\ref{eq:err_metric_first_order}), for the four leader selection strategies $(i)$--$(iv)$. 

We can note the following: strategy~$(i)$ (constant leader, Fig.~\ref{fig:simulation_first_order}(b)) has clearly the worse performance in minimizing $\|\Bolde(t)\|_{k_n}$ over time, while strategies $(ii)$--$(iii)$ (local and global leader selection, Figs.~\ref{fig:simulation_first_order}(c--d)) are able to quickly minimize $\|\Bolde(t)\|_{k_n}$ thanks to a suitable leader choice at every $T$. Interestingly, the performance of both strategies is almost the same (although strategy $(iii)$ performs slightly better): this indicates that the \emph{locality} of Algorithm~\ref{algo:1} (choosing the next leader only within the set $\calL_{k-1}$) does not pose a strong constraint, and it actually results in a less erratic leader choice (compare Fig.~\ref{fig:simulation_first_order}(c)-top with Fig.~\ref{fig:simulation_first_order}(d)-top). Finally, as one would expect, strategy~$(iv)$ (random leader selection) performs better than strategy~$(i)$ but convergence time is much worse than the other optimization strategies, being roughly $4.2$ times the convergence time of strategies $(ii)$--$(iii)$ ($\sim3$\,s vs. $\sim0.7$\,s, respectively), thus confirming the effectiveness of an \emph{active} leader selection w.r.t.~a random one.

\begin{figure*}[t]
\centering
\subfloat[]{\includegraphics[width=0.33\textwidth]{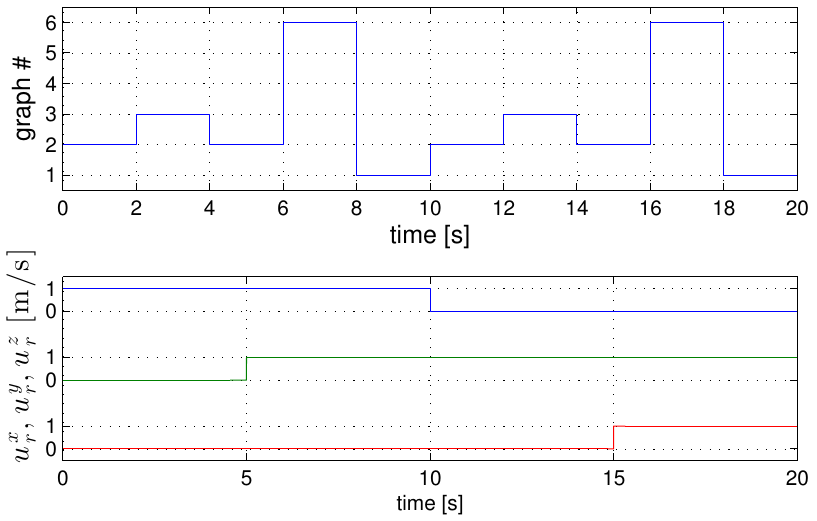}}
\subfloat[]{\includegraphics[width=0.33\textwidth]{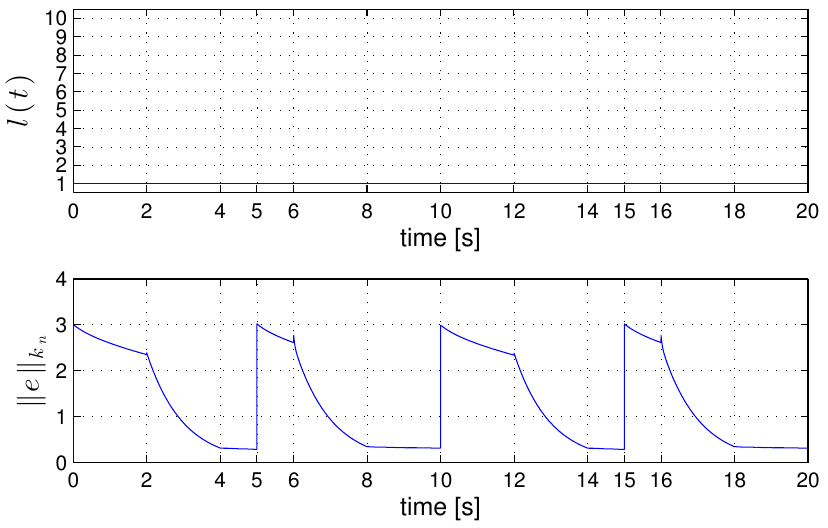}}\hfill
\subfloat[]{\includegraphics[width=0.33\textwidth]{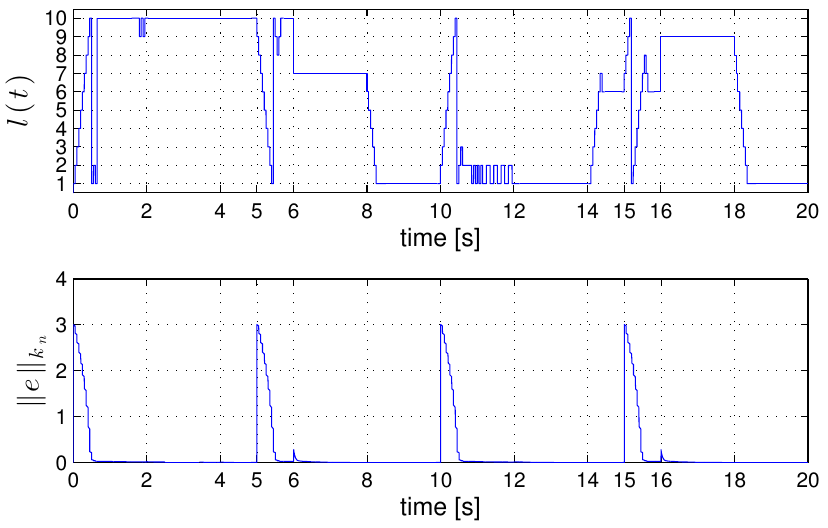}}\\
\hfill\subfloat[]{\includegraphics[width=0.33\textwidth]{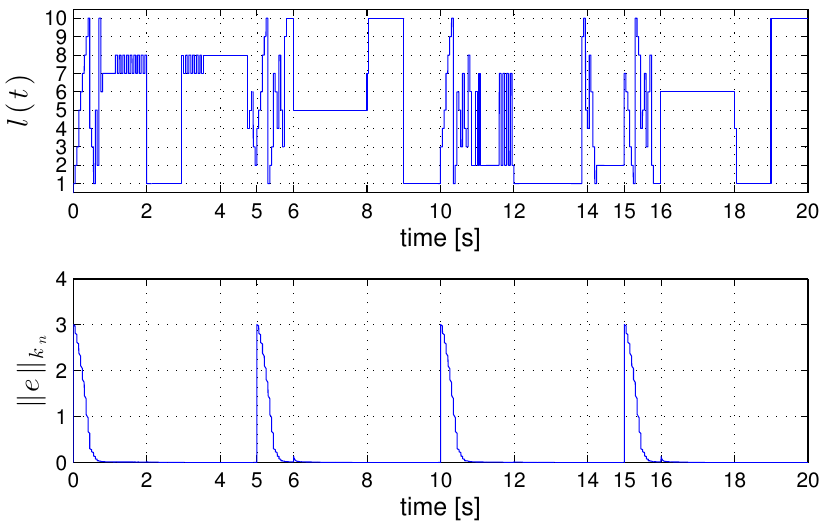}}\hfill
\subfloat[]{\includegraphics[width=0.33\textwidth]{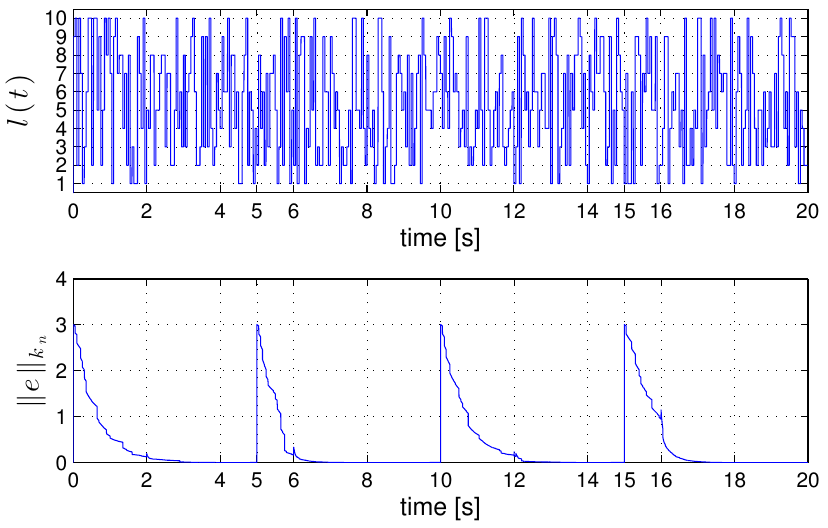}}\hfill~
\caption{Results of the four simulation runs for the first-order leader selection. Fig.~\ref{fig:simulation_first_order}(a)-top reports the current graph $\calG$ topology with the indexing defined in Fig.~\ref{fig:canonical_graphs}, and Fig.~\ref{fig:simulation_first_order}(a)-bottom shows the three components of the piece-wise constant reference velocity $\Boldu_r(t)$. Figures~\ref{fig:simulation_first_order}(b--e) then depict the identity of the current leader $l(t)$ and the error metric $\|\Bolde(t)\|_{k_n}$ for the four leader selection strategies considered in the simulations, i.e., constant leader, local leader selection, global leader selection and random leader selection, respectively. Note how the constant leader selection has the worst performance in minimizing $\|\Bolde(t)\|_{k_n}$ (Fig.~\ref{fig:simulation_first_order}(b)), followed by the random leader selection case (Fig.~\ref{fig:simulation_first_order}(e)). The local and global leader selection cases are instead able to quickly minimize $\|\Bolde(t)\|_{k_n}$ with a comparable performance.}
\label{fig:simulation_first_order}
\end{figure*}

\section{Conclusions And Future Works}\label{sec:concl}

This paper addresses the problem of \emph{online leader selection} for a group of agents in a leader-follower scenario: the identity of the leader is left free to change over time in order to optimize the performance in tracking an external velocity reference signal and in achieving a desired formation shape. The problem is solved by defining a suitable \emph{tracking error metric} able to capture the effect of a leader change in the group performance. Based on this metric, an online and decentralized leader selection algorithm is then proposed, which is able to persistently select the best leader during the agent motion. The reported simulation results clearly show the benefits of the proposed strategy when compared to other possibilities such as keeping a constant leader over time (as typically done), or relying on a random choice.

As future developments we want consider the possibility of developing similar results for the second-order case (we already have some preliminary results for a particular choice of the control gains). We also want to extend our analysis to the case of multiple masters/leaders. Finally, it will be also worth to consider decentralized online leader selection schemes for other optimization criteria such as, e.g., controllability. 

\bibliographystyle{unsrt}        %
\bibliography{bibAlias,bibMain,bibNew,bibAF}           %

\end{document}